\documentclass[sn-mathphys,Numbered,twocoloumn]{sn-jnl}

\usepackage{graphicx}%
\usepackage{multirow}%
\usepackage{amsmath,amssymb,amsfonts}%
\usepackage{amsthm}%
\usepackage{mathrsfs}%
\usepackage[title]{appendix}%
\usepackage{xcolor}%
\usepackage{textcomp}%
\usepackage{manyfoot}%
\usepackage{booktabs}%
\usepackage{algorithm}%
\usepackage{algorithmicx}%
\usepackage{algpseudocode}%
\usepackage{listings}%
\usepackage[normalem]{ulem}

\theoremstyle{thmstyleone}%
\theoremstyle{thmstyletwo}%

\theoremstyle{thmstylethree}%

\begin{document}
	
	\newcommand{\gs}[1]{\textcolor{purple}{\it #1}}
		\newcommand{\shub}[1]{\textcolor{blue}{\it #1}}
	\newcommand{\gsout}[1]{\textcolor{purple}{\sout{#1}}}
	\newcommand{\sh}[1]{\textcolor{magenta}{\it #1}}
	\newcommand{\shout}[1]{\textcolor{magenta}{\sout{#1}}}

\title{Radiative neutron capture rate of $^{11}$B$(n,\gamma)^{12}$B reaction from the Coulomb dissociation of $^{12}$B}


\author[1]{\fnm{} \sur{Shubhchintak}}\email{shubhchintak\_phy@pbi.ac.in}
\equalcont{These authors contributed equally to this work.}
\author*[2]{\fnm{G.} \sur{Singh}}\email{gsingh@uni-mainz.de}
\equalcont{These authors contributed equally to this work.}
\author[3]{\fnm{R.} \sur{Chatterjee}}\email{rchatterjee@ph.iitr.ac.in}
\author[4]{\fnm{M.} \sur{Dan}}\email{mdan@rpr.amity.edu}

\affil[1]{Department of Physics, Punjabi University Patiala, Patiala  147002, Punjab, India}

\affil*[2]{Institut f{\"{u}}r Kernphysik, Johannes Gutenberg-Universit\"{a}t Mainz, 55099, Mainz, Germany}
\affil[3]{Department of Physics, Indian Institute of Technology Roorkee, Roorkee, 247667, Uttarakhand, India}
\affil[4]{Amity School of Engineering and Technology, Amity University Chhattisgarh, Raipur, 493225, India}


\abstract{We calculate the $^{11}$B$(n,\gamma)^{12}$B reaction rate which is an important constituent in nucleosynthesis networks and is contributed by resonant as well as non-resonant capture. For the resonant rate, we use the narrow resonance approximation whereas the non-resonant contribution is calculated with the Coulomb dissociation method for which we use finite-range distorted wave Born approximation theory. We then compare our calculated rate of $^{11}$B$(n,\gamma)^{12}$B reaction with those reported earlier and with other charged particle reactions on $^{11}$B.}

\keywords{radiative capture reactions, neutron capture, Coulomb breakup, exotic nuclei, indirect techniques}



\maketitle

\section{Introduction}\label{sec1}

{Exciting and innovative recent upgrades in experimental facilities all over the world has meant that there has been an unprecedented interest in the study of exotic nuclei. Most of these lie away from the valley of stability and closer to the drip line. In fact, ever since the discovery of halos in various nuclei \cite{TSK13PPNP,THH85PRL,THH85PLB,OBC01NPA,BKT20PRL}, the amount of research near the drip lines has skyrocketed. Part of the interest in the nuclei of these regions is also due to the key astrophysical implications they have, the most vital of which are the nucleosynthesis and abundance predictions \cite{TSK01AAS,CD20JAA,KM17RPP,BDR21PRC,GORIELY98PLB,CGH97PRL,AT22AAR}.}
The lighter nuclei, namely H, He and Li, were produced mainly during the Big Bang Nucleosynthesis \cite{AT22AAR}, but an understanding of the production of metallic nuclei presents a non-trivial challenge.
Since the neutron is Coulomb neutral, naturally the number of isotopes of any given nucleus are more on the neutron rich side than the proton rich side\footnote{This is also a reason that theoretically, the addition of neutrons has proven to be easier to analyse and study than the protons.}. More than half of these neutron rich nuclei, especially those heavier than iron, are made through processes, the seeds of which are known to pass through their lighter, stable and neutron rich exotic siblings \cite{TSK01AAS,SKM05AAS,CD20JAA,BSC23PRC,AT22AAR}.
One of the most important of these lighter nuclei, not only from the point of view of life on Earth, but also the production of heavier elements, is Carbon. Various networks and series of reactions have been predicted to lead to Carbon isotopes \cite{Kajino90AJ,Bert10AIP,Liu01PRC,Dubo20AP,Lee10PRC}, which then are pathways to higher charge and/or mass isotopes, for example\footnote{Of course, there are other networks and chains that lead to Carbon production and utilisation in nucleosynthesis like the CNO cycles \cite{WGU10ARNP}.},

\noindent
... $^4$He$(t,\gamma)^7$Li$(n,\gamma) ^8$Li$(\alpha,n)$\textbf{$^{11}$B$(n,\gamma)^{12}$B}$(\beta)^{12}$C$(n,\gamma)^{13}$C$(n,\gamma)^{14}$C...,\\
\noindent
...  $^4$He$(t,\gamma)^ 7$Li$(n,\gamma)^ 8$Li$(n,\gamma)^ 9$Li$(\beta)^ 9$Be$(n,\gamma)^{10}$Be$(\beta)^{10}$B$(n,\gamma)$\textbf{$^{11}$B$(n,\gamma)^{12}$B}$(\beta)^{12}$C...,\\
\noindent
... $^4$He$(t,\gamma)^ 7$Li$(t,n)^9$Be$(n,\gamma)^{10}$Be$(\beta)^{10}$B$(n, \gamma)$\textbf{$^{11}$B$(n,\gamma)^{12}$B}$(\beta)^{12}$C$(n,\gamma)^{13}$C$(n,\gamma)^{14}$C. .

As seen above, a key reaction in all these chains is the $^{11}$B$(n,\gamma)^{12}$B radiative neutron capture reaction. It is also an important reaction for evaluation of primitive CNO mass fraction \cite{Coc12AJ}. Its rate, however, is not very precisely known although there have been a few efforts \cite{Dubo20AP,Lee10PRC,Liu01PRC,Lin03PRC}. This is partly due to the paucity in the data available for the structure and reaction variables for $^{12}$B. Precise values of nuclear structure and reaction observables are indispensable in such reaction rate calculations, as small variations in their numbers can develop vastly different results, thus affecting the overall predicted abundances. It is essential not only from the point of view of experiments, but also theory to constrain these uncertainties in nuclear variables, especially for nuclei away from the valley of stability, where our knowledge of nuclear phenomena is rather limited at present.
Moreover, carrying out direct reactions at the pertinent temperatures available in the stellar plasma for such rate computations is extremely tedious, as with the present day experimental techniques, it is difficult to access those low energy regimes. Hence, indirect approaches have been developed that provide frameworks to describe the reactions in alternate manners. 


While a variety of indirect methods are popular \cite{CRA14PRC,BBR86NPA,BSM16JPc,TBC14RPP,CRA16PRC,SLL14PRL}, Coulomb dissociation (CD) is one of the tools available where the reaction observables computed for the kinematically reverse reaction are used to find the final rates of the reaction concerned. The method employs the virtual photon field of a Coulomb heavy target to photodissociate the projectile and then study the time reversed capture reaction by using the principle of detailed balance \cite{BBR86NPA}. For our implementation, we use CD through the vision of the finite-range distorted wave Born approximation (FRDWBA) theory in its post form. The upshot of using the post form FRDWBA, a fully quantum mechanical perturbative approach, is that it includes the target fragment interactions to all orders. 
The ground state wave function of the projectile is the only input that goes into the transition matrix calculations, and it includes the entire non-resonant continuum \cite{SC14NPA,SSC16PRC,CBS00NPA}. 
This is an added advantage over several first-order theories \cite{WA79NPA,TB05NPA}, where fixing the continuum states can be tricky \cite{TB05NPA}.  The known spectrum of $^{12}$B contains several low-energy narrow resonances and FRDWBA will be useful to constrain the non-resonant contribution towards the reaction rate.

Once the non-resonant contribution is known, one can then add the resonant contribution for individual resonances by using the usual Breit-Wigner approximation \cite{HBG10ADNT} (given that there are low-energy non-interfering narrow resonances in the $^{11}$B$-n$ system). It is worth mentioning that for a given temperature range there will be a small energy window that can contribute to the reaction rate and therefore, the whole low energy spectrum may not be useful for the rate calculation. So, apart from demarcating the contribution of the non-resonant reaction rate we also try to see the contribution of individual resonances to the reaction rate, mainly at low temperature.
The $^{11}$B$(n,\gamma)^{12}$B reaction also competes with proton and $\alpha$ capture reactions in various astrophysical modelings, such as homogeneous and inhomogeneous Big-Bang nucleosynthesis \cite{Wang91PRC, Rauscher94AJ}. We, therefore, compare our results with the rates of the $p$ capture as well as $\alpha$ capture reactions obtained from the literature. Such a comparison could be useful for proton-Boron fusion reactors \cite{Chirkov23Plasma,HB11_23JFE}, an alternative to deuterium-tritium fusion reactors.

We organize this paper as follows. A brief description of the FRDWBA theory of breakup reactions is given in section \ref{sec2}. In section \ref{sec3}, we present our results on breakup cross-sections and subsequently, radiative capture cross-sections and rates. Section \ref{sec4} presents the conclusions of our work.

\section{Formalism}
\label{sec2}

The reaction rate for any radiative capture reaction of the form $\textit{b} + \textit{c}  \longrightarrow  \textit{a} + \gamma$ is given by,
\begin{equation}
	\begin{aligned}
		R=N_A \langle\sigma_{(n,\gamma)}\textit{v}\rangle, 
		\label{Rate}
	\end{aligned}
\end{equation}
where, $\langle\sigma_{(n,\gamma)}\textit{v}\rangle$ is the reaction rate per particle pair and is averaged over Maxwell-Boltzmann velocity distribution \cite{RolfsBook}, while $N_A$ is the famous Avogadro number. One can write $\langle\sigma_{(n,\gamma)}\textit{v}\rangle$ in units of $\textrm{cm}^3$ $\textrm{s}^{-1}$ as
\begin{eqnarray}
	\langle\sigma_{(n,\gamma)}\textit{v}\rangle &=& \sqrt{\frac{8}{\pi\mu_{bc}\,(k_BT)^3}}\int_{0}^{\infty}\sigma_{(n,\gamma)}(E_{rel})
	\times E_{rel}\exp(-E_{rel}/k_BT)dE_{rel},
	\label{MARR}
\end{eqnarray}
where, $k_B$ is the Boltzmann constant, $T$ is the temperature, $\mu_{bc}$ is the reduced mass of two fragments $b$ and $c$ and $E_{rel}$ (or $E_{c.m.}$ in some notations) is the relative or the center of mass energy of the $b-c$ system. $\sigma_{(n,\gamma)}$ is what is called the radiative capture cross-section and gives us an effective measure of the interaction probability of the particles $b$ and $c$ during a radiative capture reaction.

Nuclear astrophysics usually deals with the low energy nuclear reactions and direct measurements in many cases often become difficult. Therefore, indirect techniques have been developed over the years to study such reactions in an alternative manner \cite{BBR86NPA, CRA16PRC, TBC14RPP,NSC15PRC,BCS08PRC,BSM16JPc}. 
One such indirect method uses the virtual photon field of a heavy target to break the projectile nucleus in question into a core and its valence nucleon(s), enabling the calculation of its breakup cross-section. This breakup cross-section can then be used to compute the cross-section of the disintegration process of the projectile due to actual $\gamma$-rays. This is called the photodisintegration cross-section. The radiative capture cross-section relates to this time reversed photodisintegration cross-section via the principle of detailed balance as:

\begin{equation}
	\begin{aligned}
		\sigma_{(n,\gamma)} = \dfrac{2(2\textit{j}_a+1)}{(2\textit{j}_b+1)(2\textit{j}_c+1)}\dfrac{\textit{k}_\gamma^2}{\textit{k}^2}{\sigma_{(\gamma,n)}},
		\label{pdbal}
	\end{aligned}
\end{equation}

where $\textit{j}_{i}$ ${(i = a, b, c)}$ is the total spin of the particle/cluster,  $\textit{k}$ $\left(=\dfrac{\sqrt{2\mu_{bc} E_{rel}}}{\hbar}\right)$ is the wave number of the relative motion between $b$ and $c$, while $\textit{k}_\gamma$ is photon wave number. 
We use this approach to calculate the non-resonant capture cross section $(\sigma_{(n,\gamma)})_{nr}$ and subsequentl,y the non-resonant rate. For the resonant rate, as mentioned earlier, we use the Breit-Wigner formula because of the low-energy narrow resonances involved in the system.

\begin{figure}[ht]
	\centering
	\centering
	\includegraphics[height=7.5cm, clip,width=0.75\textwidth]{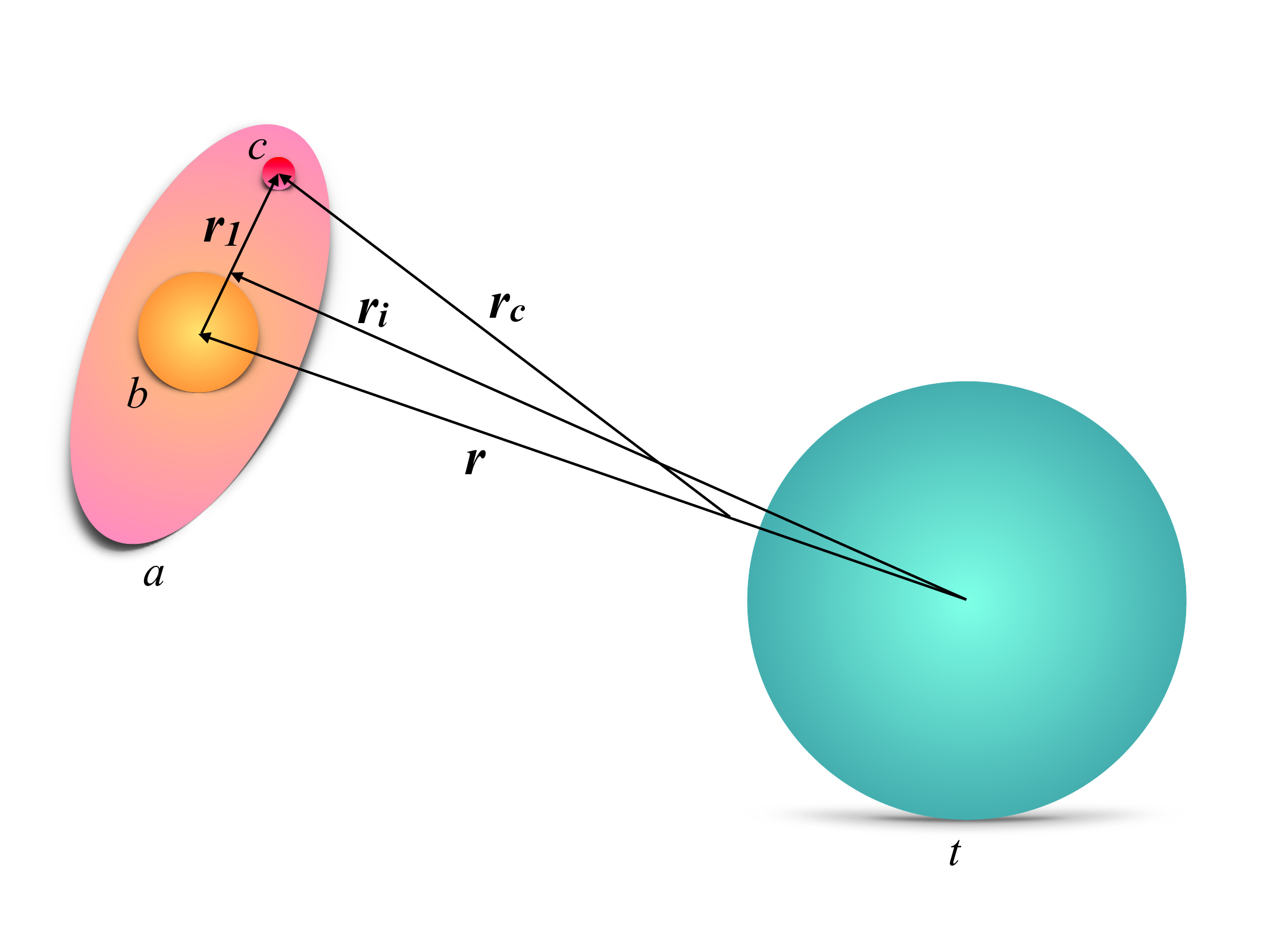}
	\caption{The Jacobi coordinate scheme for the elastic breakup reaction \textit{a} + \textit{t} $\longrightarrow$ \textit{b} + \textit{c} + \textit{t}. A heavy target, $t$, ensures a strong dynamic Coulomb field sufficient to break the lightly bound projectile, $a$.}		
	\label{fig:Jacobi}		
\end{figure}

To obtain now the breakup cross-section, we contemplate the forward reaction with a (loosely bound) two-body projectile $a$ $(= b + c)$ impinging on a heavy target $t$. The use of a heavy target ensures that its large dynamic Coulomb field felt by the incoming projectile coerces the latter to undergo elastic breakup and give off a core $b$ and a valence nucleon $c$, i.e., \textit{a} + \textit{t} $\longrightarrow$ \textit{b} + \textit{c} + \textit{t}. The coordinate scheme for the breakup reaction is displayed in Fig. \ref{fig:Jacobi}.

{The triple differential cross-section for such a reaction is given by,} 
\begin{equation}
	\begin{aligned}
		\dfrac{d^3\sigma}{dE_{b}d\Omega_{b}d\Omega_{c}} = \dfrac{2\pi}{\hbar v_{at}}\rho{(E_{b},\Omega_{b},\Omega_{c})}\sum_{l,m}|\beta_{lm}|^{2},
		\label{triple}
	\end{aligned}
\end{equation}
{where 
	$\rho{(E_{b},\Omega_{b},\Omega_{c})}$ is the three body final state phase space factor \cite{Fuchs82NIM} and $v_{at}$ is the relative velocity between the projectile and target in the initial channel.}

{Eq. (\ref{triple}) above contains the information of the transitions of the initial states to the final states due to the effect of the interactions through the reduced transition amplitudes $\beta_{lm}$. This reduced transition apmplitude, in the framework of the finite range distorted wave Born approximation theory is described as, }

\begin{eqnarray}
	\beta_{lm}(\textbf{q}_{b},\textbf{q}_{c};\textbf{q}_{a}) &=& \left\langle \chi_{b}^{(-)}(\textbf{q}_{b},\textbf{r})
	\chi_{c}^{(-)}(\textbf{q}_{c},\textbf{r}_{c})\right\vert V_{bc}(\textbf{r}_{1}) \left\vert
	\phi_{a}^{lm}(\textbf{r}_{1})\chi_{a}^{(+)}(\textbf{q}_{a},\textbf{r}_{i})\right\rangle.
\end{eqnarray}

{Here, $\phi_{a}^{lm}$ is the ground state wave function of the projectile \textit{a} such that it is the eigenfunction of the Hamiltonian that includes the two-body potential  $V_{bc}$, which for our purpose is taken of the Woods-Saxon type and can include possible deformation \cite{SSC16PRC,SC14NPA}. $l$ is the orbital angular momentum and $m$ as its projection, while $\textbf{q}_{i}$ is the wave vector in the Jacobi co-ordinate system. The $(-)$ and the $(+)$ represent the incoming and outgoing wave boundary conditions, whereas $\chi_i$ is the pure Coulomb distorted wave of the corresponding particle. The only input to the theory, as is evident, is the ground state three-body wave function.
	
	Now, for a neutron rich projectile breaking into a core and a valence neutron, the distorted wave for particle $c$ expectedly gets replaced by a plane wave, which is then convoluted with other distorted waves in the above integral. This reduction in the number of distorted waves is advantageous in that it significantly enhances the convergence of the integral. 
	
	One can then obtain the relative energy spectrum \cite{BCS08PRC,CS18PPNP} of the outgiong fragments by multiplying with the appropriate Jacobian and integrating over the appropriate angles: 
	
	\begin{equation}
		\begin{aligned}
			\dfrac{d\sigma}{dE_{rel}}\ = 
			\dfrac{\mu_{bc}\mu_{at} p_{bc}p_{at}}{{(2\pi)^5\hbar^7}\textit{v}_{at}}
			\int\sum_{l,m}|\beta_{lm}|^{2}\dfrac{1}{2l+1}{d\Omega_{bc}d\Omega_{at}}.
			\label{rel-eng}
		\end{aligned}
	\end{equation}
	
	{For a pure Coulomb breakup, i.e., if the nuclear effects are small and negligible, one may also relate the relative energy spectrum to the photodisintegration cross-section via \cite{BBR86NPA},}
	
	\begin{equation}
		\begin{aligned}
			\dfrac{d\sigma}{dE_{rel}} = \dfrac{1}{E_{\gamma}}\sum_{\pi,\lambda}{\sigma^{\pi\lambda}_{(\gamma,n)}}{n_{\pi\lambda}} ,
			\label{rel-photo}
		\end{aligned}
	\end{equation}
	{where $n_{\pi\lambda}$ is the virtual photon number, with $\pi$ corresponding to electric or magnetic transitions and $\lambda$ being the multipolarity. The $\gamma$ - energy is given by $E_{\gamma}= E_{rel} + S_n$, where $S_n$ is the one neutron separation energy of the projectile nucleus. It is now essential that the electromagnetic transitions be dominated by a single multipolarity and type. Considering the cases where this is true, the summation in Eq. (\ref{rel-photo}) vanishes leaving only the term corresponding to the multipolarity of the transition. Then for a reaction dominated by an electric transition of order $\lambda$, the photodisintegration cross-section can be obtained via,} 
	
	\begin{equation}
		\sigma_{(\gamma,n)} = \left(\dfrac{E_\gamma}{n_{E\lambda}}\right)\dfrac{d\sigma}{dE_{rel}},
		\label{photo}
	\end{equation}
{with $n_{E\lambda}$ being the virtual photon number for the $E\lambda$ transition. Once the photodisintegration cross-section is known, the non-resonant reaction rate of the radiative capture reaction can be computed through the implementation of Eqs. (\ref{pdbal}), (\ref{MARR}) and (\ref{Rate}), respectively. For more details, one is referred to \cite{SCS17PRC,SSC17PRC,DSC19PRC,NSC15PRC}, where this method of obtaining the reaction rate via the FRDWBA theory has been successfully used to calculate and analyse several reactions of astrophysical interest.}

\section{Results and discussions}
\label{sec3}

$^{12}$B is an odd-odd nucleus with $N$ = 7 and $T_{1/2}$ = 20.20\,ms. Its ground state spin-parity  ($J^{\pi}$) is $1^+$ which is obtained by coupling the  $^{11}$B core in the ground state ($3/2^-$) with a $1p_{1/2}$ neutron. 
\begin{figure}[ht]
\centering
\centering
\includegraphics[height=7.5cm, clip,width=0.65\textwidth]{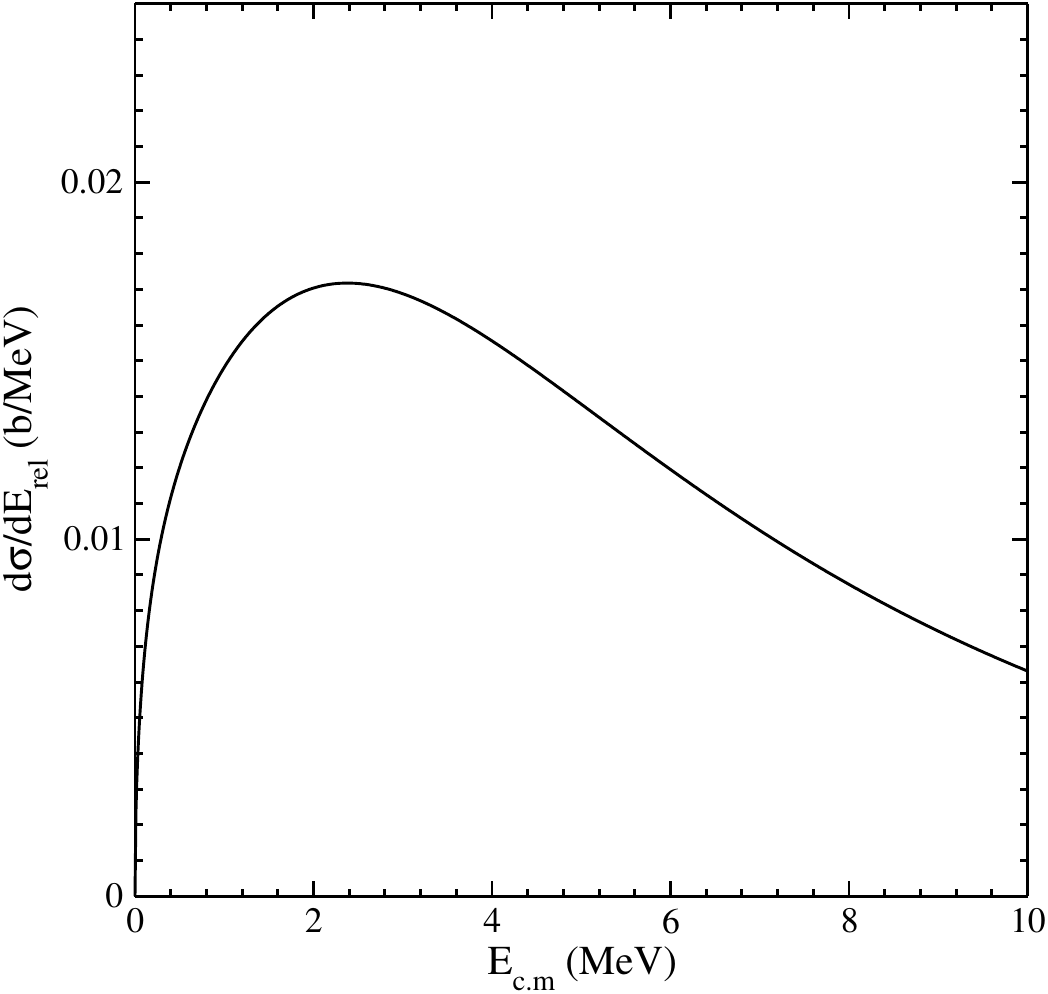}
\caption{Relative energy spectrum in the Coulomb breakup of $^{12}$B on $^{208}$Pb calculated at beam energy of 60 MeV/nucleon.}		
\label{fig:rel_eng}		
\end{figure}		
The one neutron separation energy is $S_n$ = 3.3696(13)\,MeV \cite{NNDC}.

We calculate the ground state wave function of $^{12}$B by fixing the depth of a Woods-Saxon potential to reproduce the $S_n$. Radius and diffuseness parameters of the potential are fixed as 1.35 fm and 0.65 fm \cite{Bely18PRC}, respectively. The wave function calculated in this way is then used as an input to calculate the relative energy spectra [Eq. (\ref{rel-eng})], which are then further required to calculate the non-resonant capture cross-sections of the  $^{11}$B$(n,\gamma)^{12}$B radiative capture reaction. 
	
We contemplate the elastic breakup of $^{12}$B into $^{11}$B core and a neutron in the Coulomb field of a Pb target at a beam energy of 60 MeV/nucleon. The relative energy spectrum for the reaction is shown in Fig. \ref{fig:rel_eng}. 
Due to the non-availability of the corresponding experimental data, to which calculations can be normalised, we adopt the latest spectroscopic factor of 0.69 taken from Ref. \cite{Bely18PRC}. The same value was obtained in Ref. \cite {Lee10PRC}. Nevertheless, CD data will be important to verify and constrain the limits on the non-resonant cross section of $^{11}$B$(n,\gamma)^{12}$B reaction when calculated using the current approach. 	

As in Ref. \cite{HBG10ADNT}, we also consider $E1$ transitions from the continuum to the ground state of $^{12}$B. They are, indeed, the dominating transitions. Eventually, by using Eq. (\ref{photo}), photo-disintegration cross-sections are then calculated. Non-resonant capture cross-sections for $^{11}$B$(n,\gamma)^{12}$B reaction are then easily obtained from the photo-disintegration cross-section by using the principle of detailed balance. 

In Fig. \ref{fig:cap}, we plot our non-resonant capture cross sections and compare these with the experimental data from Ref. \cite{Imhof62PR}. 
	\begin{figure}[ht]
		\centering
		\includegraphics[height=7.5cm, clip,width=0.65\textwidth]{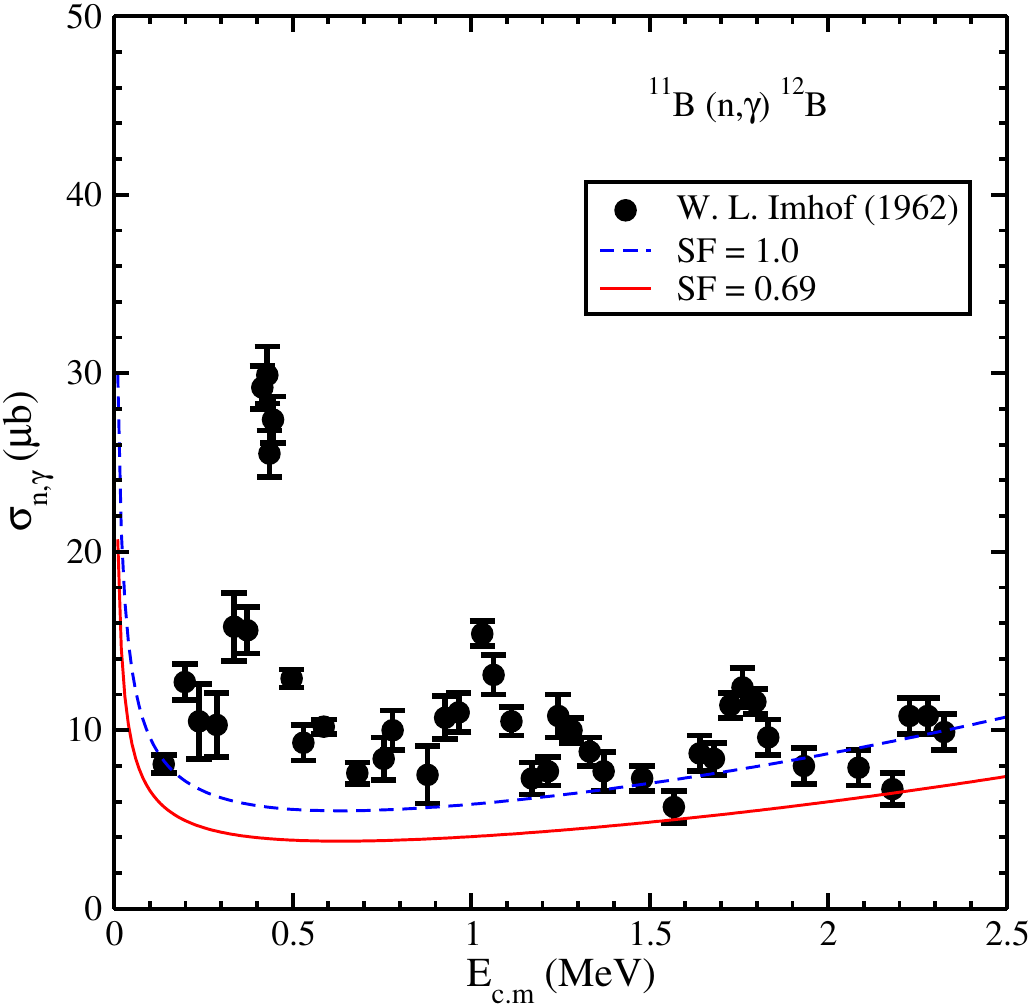}
		\caption {The $^{11}$B$(n,\gamma)^{12}$B radiative capture cross-section as a function of the relative energy $E_{c.m.}$. Dashed and solid lines are calculations using two different spectroscopic factors of 1 and 0.69, respectively.  Experimental data are taken from Ref. \cite{Imhof62PR} for the direct capture. For more details, see text.}
		\label{fig:cap}
	\end{figure}	
Dashed and solid lines represent the calculations with spectroscopic factors 1 and 0.69, respectively. Our calculated cross-sections behave similar to those reported in Ref. \cite{HBG10ADNT}, which were calculated using a potential model. However, for energies $< 500$ keV, our cross-sections are smaller than those calculated by the potential model. 
As discussed earlier and as is also clear from the data that in this case, the low energy capture cross-sections are contributed by several narrow resonances [at excitation energies ($E_x$) of 3.3889, 3.764, 4.311, 4.54 and 5.00 MeV, respectively]. The total cross-section can then be explained as a sum of the non-resonant and the resonant cross-sections \cite{HBG10ADNT}. However, it will be interesting to see which of these resonances actually contribute to the rate. 

	\begin{figure}[ht]
		\centering
		\includegraphics[height=7.5cm, clip,width=0.65\textwidth]{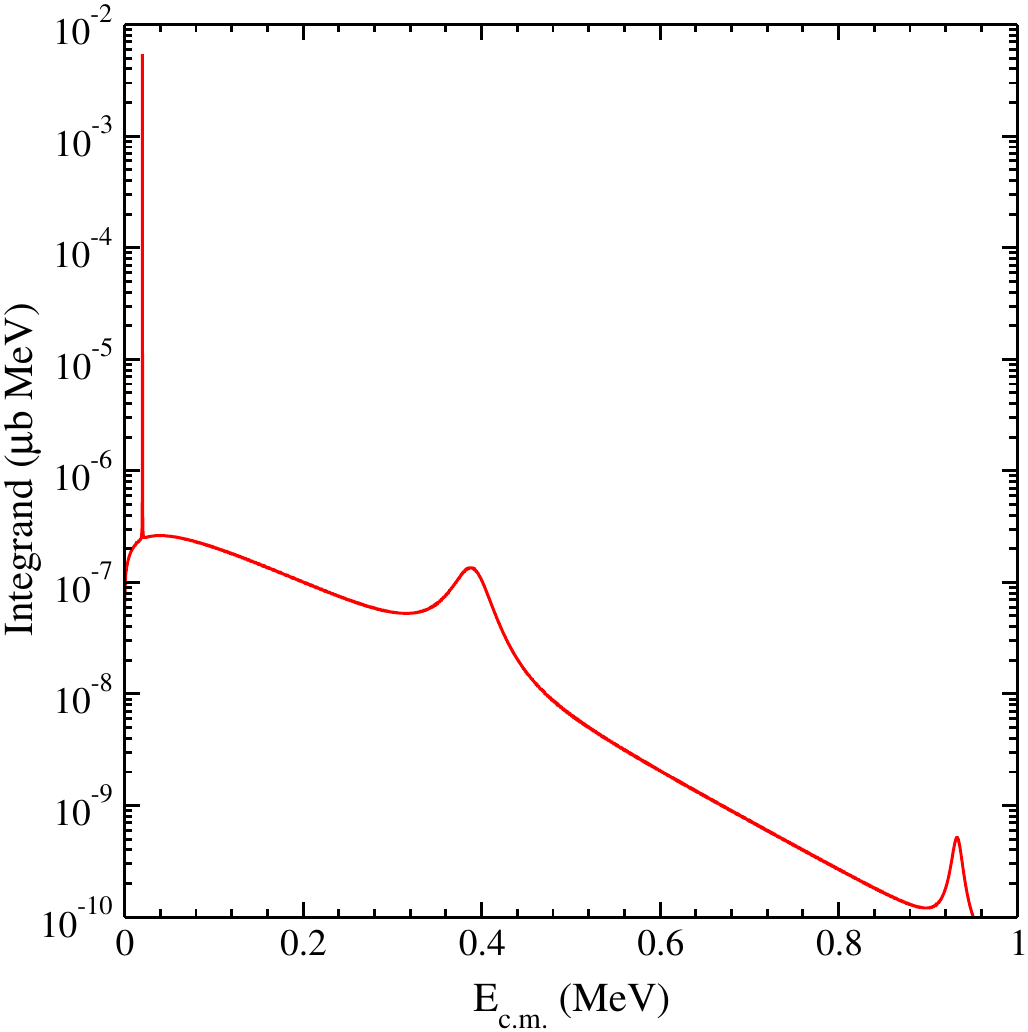}
		\caption {Integrand in Eq. (\ref{MARR}), plotted as a function of relative energy ($E_{c.m.}$ or $E_{bc}$) for $T_9 = 1$ . For more details see text.}
		\label{fig:integ}
	\end{figure}	

To check for this, in Fig. \ref{fig:integ} we plot the integrand in Eq. (\ref{MARR}) as a function of relative energy of the  $^{11}{\rm B}-n$ at $T_9 = 1$, which is the temperature in units of $10^9$ K. $\sigma_{(n, \gamma)}$ used in the equation includes a sum of the non-resonant as well as resonant cross-sections. The resonant cross-section, in turn, consists of a sum of the first 5 low lying energy resonances calculated using the Breit-Wigner formula, where the resonance parameters are adopted from Refs. \cite{Imhof62PR,Kelley17NPA}. It is abundantly clear from the figure that even at $T_9 = 1$, which is roughly equal to a c.m. energy of about 0.1\,MeV, the dominant contribution to the reaction rates comes from the resonance at 20.8\,keV, which is a narrow resonance with $\Gamma_{\gamma} = 25 \pm 8$ meV, $\Gamma_n = 3.1 \pm 0.6$ eV \cite{Mooring69PR}. All the other resonances contributue negligibly in comparison to this resonance, which must be modelled properly. Therefore, it is adequate to consider only this resonance and one does not need the entire resonant spectrum of $^{12}$B to calculate the reaction rates, especially at low temperature. Indeed, this was also pointed out in Ref. \cite{Lee10PRC}.


Knowing that only one resonance is significant in this case and is narrow, we see that the reaction rate per mole for a narrow resonance can be approximated as \cite{Iliadis}:
\begin{eqnarray}
N_A\langle\sigma v\rangle_r = 1.54 \times 10^{11} (\mu_{bc} T_9)^{-3/2} \times (\omega \gamma) \, exp\Big(\frac{-11.605 E_r}{T_9}\Big),
\end{eqnarray}
where $E_r$ is the resonance energy in c.m. and $\omega \gamma$ is the resonance strength in units of MeV.
The value of $\omega \gamma$ was reported to be $(2.2 \pm 0.8) \times 10^{-2}$ eV in Ref. \cite{Rauscher94AJ}, whereas the latest value reported in Ref. \cite{Lee10PRC} is $(3.3 \pm 0.6) \times 10^{-2}$ eV, which is around 50\% larger and results in about a 50\% increase in the reaction rate \cite{Lee10PRC}. We also use this value of $\omega \gamma$ for resonant rate calculations. However, our non-resonant rates are significantly larger than those reported in Ref. \cite{Lee10PRC}, which results in an increase in the sum total rate. It would, therefore, be interesting to compare our total ($n, \gamma$) reaction rates with those reported earlier in the literature as well as with those of other charged particle reactions on $^{11}$B. 
	\begin{figure}[ht]
		\centering
		\centering
		\includegraphics[height=7.5cm, clip,width=0.65\textwidth]{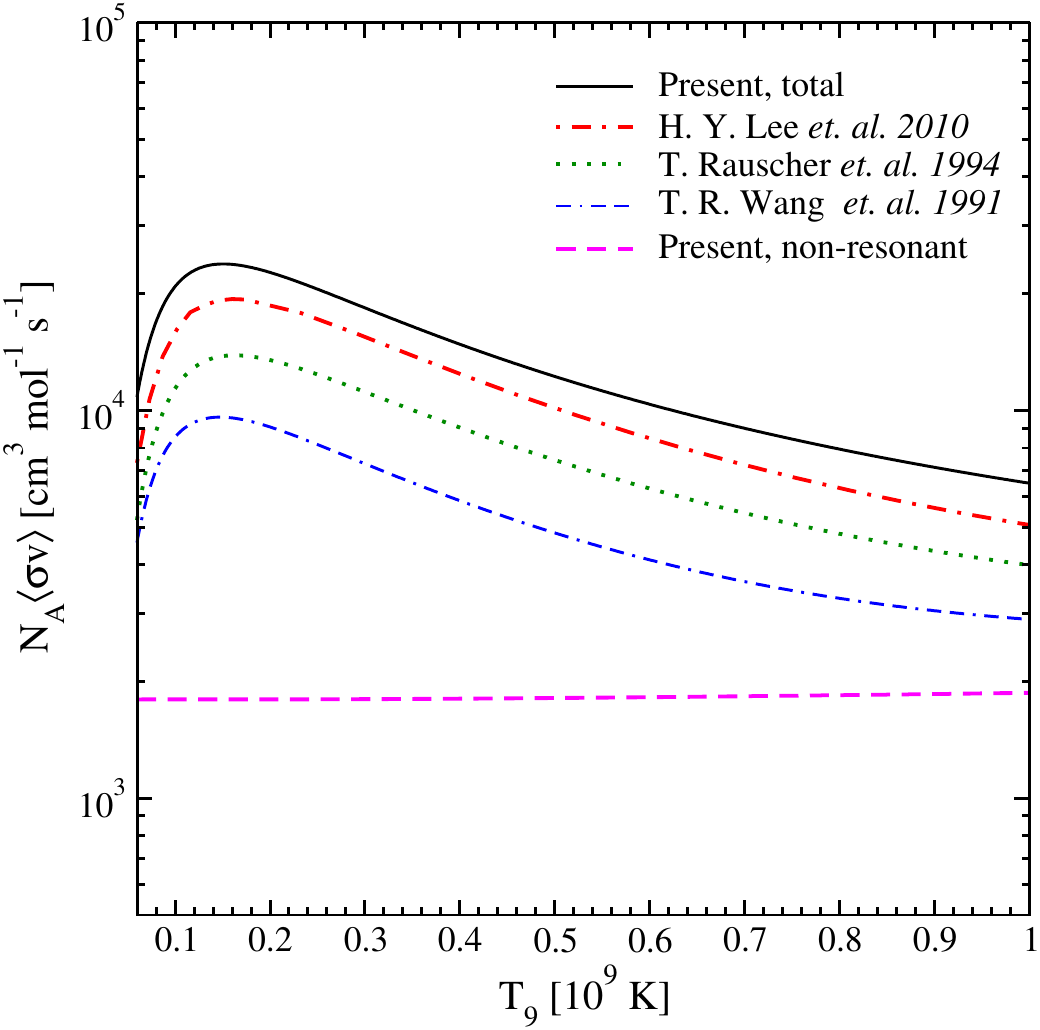}
		\caption{$^{11}$B(n, $\gamma$)$^{12}$B reaction rate (solid line) as a function of $T_9$, the non-resonant contribution is shown by dashed line. Dot-dashed, dotted and dashed-dashed-dotted lines represent rates taken from Lee \textit{et al.} \cite{Lee10PRC}, Rauscher \textit{et al.} \cite{Rauscher94AJ} and Wang \textit{et al.} \cite{Wang91PRC}, respectively. }		
		\label{fig:rate3}	
	\end{figure}	

In Fig. \ref{fig:rate3}, we display the reaction rate of the $^{11}$B(n, $\gamma$)$^{12}$B radiative capture reaction as a function of temperature ($T_9$) and compare it with some of the earlier reported works. The dot-dashed, dotted  and dashed-dashed-dotted lines represent rates reported in Refs. \cite{Lee10PRC}, \cite{Rauscher94AJ} and \cite{Wang91PRC}, respectively.
It is clear that the non-resonant rate (dashed-line) has a significant contribution to the sum which results in larger total reaction rate as compared to that reported in Ref. \cite{Lee10PRC}, from where we have adopted the resonance strength value of the 20.8 keV resonance. Note that in Ref. \cite{Lee10PRC}, non-resonant rates were calculated using capture cross-sections obtained from a potential model where the continuum potential was fixed to reproduce the experimental thermal cross section of 5.5 mb reported in Refs. \cite{Imhof62PR, MughaBook}. However, a recently proclaimed
value for the same is $9.09 \pm 0.10$ mb \cite{Fire16PRC}, that may lead to changes in the non-resonant cross section reported in Ref. \cite{Lee10PRC}.
Therefore, it is important to have some experimental findings for direct capture (non-resonant) cross section. 

This increase in reaction rate can influence the abundance of $^{12}$B and of the subsequent nuclei formed by different reactions involving $^{11}$B and $^{12}$B in various astrophysical as well as man-made scenarios. 
As mentioned, the $^{11}$B(n, $\gamma$)$^{12}$B reaction is also viewed to be important in the inhomogeneous Big-Bang model where it can compete with $\alpha$ capture reaction \cite{Wang91PRC, Rauscher94AJ} when considering the formation of heavier nuclei. On the other hand, in a standard Big-Bang scenario, a proton induced reaction on $^{11}$B, $i.e.$, the $^{11}$B(p, 2$\alpha$)$^{4}$He reaction, prevents the formation of heavier nuclei by destroying $^{11}$B to form 3 $\alpha$'s, rendering it unavailable for neutron capture. 
In fact, recent studies involving proton-Boron fusion reactors, which are seen as an alluring alternative to the traditional Hydrogen based (deuterium-tritium) fusion reactors are mainly based on this reaction\footnote{See, for example Refs. \cite{Chirkov23Plasma,HB11_23JFE}, where the crucial purpose of proton capture reactions on $^{11}$B are discussed.}. The resulting $\alpha$'s from the $^{11}$B(p, 2$\alpha$)$^{4}$He reaction can lead to the $\alpha$ induced $^{11}$B($\alpha$,n)$^{14}$N reactions \cite{Borg23PRC}. The generated neutrons from this reaction can then, in turn, induce the $^{11}$B(n, $\gamma$)$^{12}$B further, though the resulting fractions and hence, the probabilities for this are expected to be very small \cite{HB11_23JFE}.
Therefore, it becomes important to compare rates of these different type of reactions on $^{11}$B.

	\begin{figure}[ht]		
		 \centering
		\includegraphics[height=7.5cm, clip,width=0.65\textwidth]{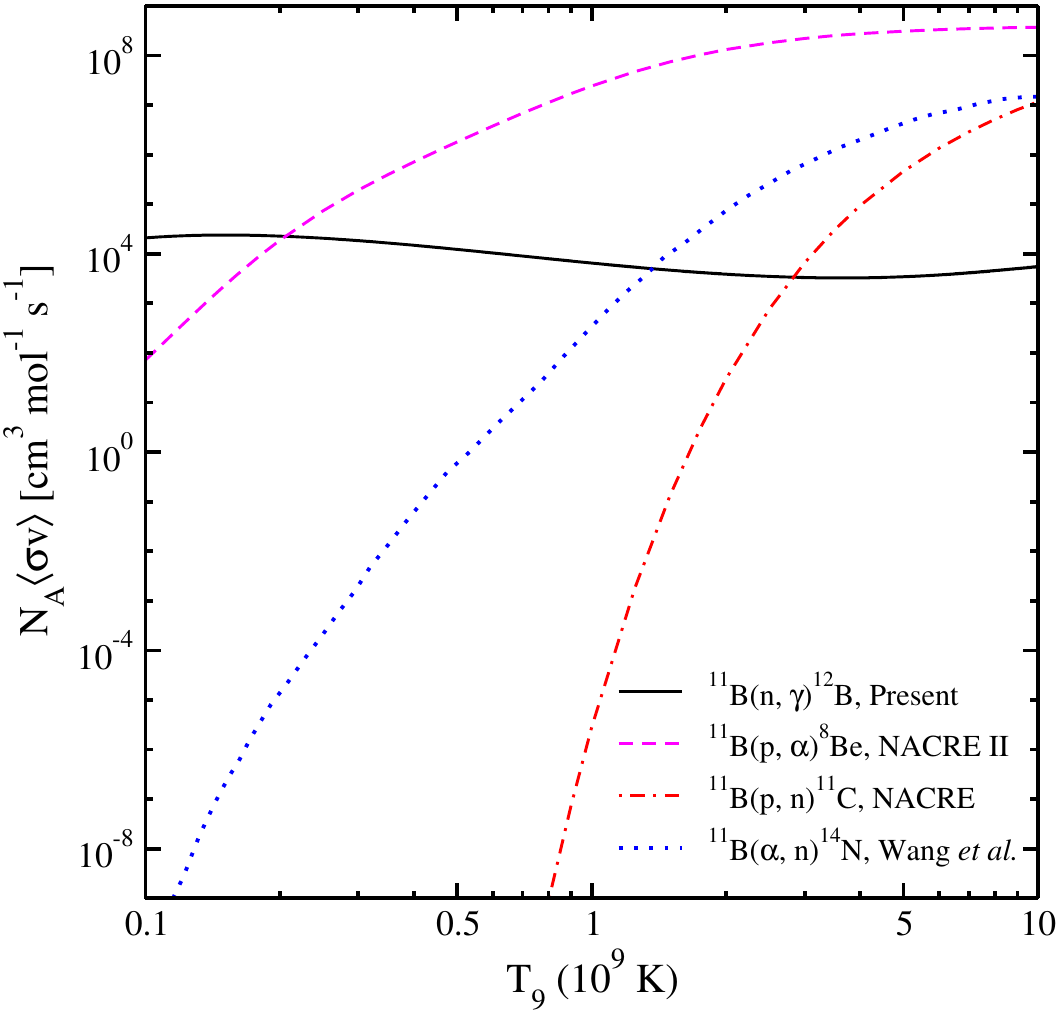}
		\caption{Comparison of $^{11}$B$(n,\gamma)^{12}$B reaction rate with various charged particle capture rates available in the literature (taken from Refs. \cite{NACREII,NACRE,Wang91PRC}). The black solid line corresponds to the FRDWBA calculations of this work. The $\alpha$-capture rates seem to be expectedly dominated by the $n$-capture rates at lower temperatures. For more details, see text.}
		\label{fig:rate_comp}			
	\end{figure}

Fig. \ref{fig:rate_comp}, bears a comparison of our newly calculated $^{11}$B(n, $\gamma$)$^{12}$B rate (solid line) with that of $^{11}$B($\alpha$, n)$^{14}$N obtained from Ref. \cite {Wang91PRC} (dotted line), $^{11}$B(p, 2$\alpha$)$^{4}$He [also represented as $^{11}$B(p, $\alpha$)$^{8}$Be]  taken from Ref. \cite{NACREII} (dashed line) and $^{11}$B(p, n)$^{11}$C from Ref. \cite{NACRE} (dot-dashed line) in the temperature range $0.1\le T_9 \le 10$. The temperature range selected here is relevant for the inhomogeneous Big-Bang model ($T_9=0.2-1.2$) and the Boron fusion reactors \cite{HB11_23JFE}. It is clear from the figure that the $^{11}$B(n, $\gamma$)$^{12}$B reaction dominates the $^{11}$B($\alpha$, n)$^{14}$N upto $T_9 \simeq 1.35$, which is slightly higher than the  $T_9 = 1.2$ declared in Ref. \cite{Wang91PRC}. This is due to our larger resonant as well as non-resonant rates as compared to Wang \textit{et al.} \cite{Wang91PRC}. A recent study \cite{Borg23PRC} also reports the $^{11}$B($\alpha$, n)$^{14}$N reaction rate by remeasuring the low energy cross-sections. However, in the temperature range $T_9 \ge 1$, their rates are nearly the same as those of Ref. \cite{Wang91PRC}. It is also clear that in the proton rich environment, the $^{11}$B(p, 2$\alpha$)$^{4}$He dominates all the above reactions and its rate overcomes the $(n, \gamma)$ rate at relatively lower temperatures ($T_9 \ge 0.2$) as compared to other competing reactions. Note that the $^{11}$B(p, 2$\alpha$)$^{4}$He reaction is considered as the main reaction for the Boron-fusion reactors  \cite{Chirkov23Plasma}. This suggests that to safely overcome the competition with the $^{11}$B(n, $\gamma$)$^{12}$B reaction, the proton-Boron fusion reactors must achieve a temperature range above $T_9 \simeq 0.3$. As discussed in Ref. \cite{HB11_23JFE}, the $^{11}$B($\alpha$, n)$^{14}$N and $^{11}$B(p, n)$^{11}$C can be two possible neutron generator reactions in such reactors pushing for the formation of $^{12}$B again, but with very small probability for $T_9 \leq 1$. The rate comparison in Fig. \ref{fig:rate_comp} also emphasises the same.


\section{Conclusions}
\label{sec4}
An accurate prediction of nucleosynthesis paths can be done from a precise knowledge of the different stellar burning stages. Apart from the comprehension of the various hadronic interactions going on in the stellar plasma, this necessitates a clear understanding of the reaction rates (neutron capture, photodisintegration, proton and alpha capture, beta decay, etc.) involving not only heavy, but also the light and medium mass nuclei. These nuclei are important, while conjecturing various reaction networks, as it is known that varying their nuclear structure information (viz., level density, separation energy, spin-parity, etc.) while traversing along an isotopic chain can leave a significant mark on the final abundances of nuclei \cite{Mumpower16PPNP,Goriely97AA,Goriely05EPJA}.
	
Nevertheless, performing direct experiments at the energy ranges relevant to the stellar plasma is indeed very difficult, and hence, indirect techniques are used to explore the regime. Coulomb dissociation, our indirect method using the FRDWBA theory, has been successfully used in the past to compute radiative neutron capture rates for exotic nuclei in astrophysical environments \cite{SSC17PRC,SCS17PRC,BCS08PRC,CS18PPNP,NSC15PRC,DSC19PRC,Singh20JPc}. 


In the present work, we calculated the direct capture cross-section and subsequent non-resonant rate of $^{11}$B(n, $\gamma$)$^{12}$B reaction using the Coulomb dissociation of $^{12}$B on Pb in the FRDWBA framework. Resonant rate at low temperature is found to be mainly contributed by the state at 3.3889 MeV, which is modelled by using the Breit-Wigner approximation. Our calculations reveal that the non-resonant rate has a significant contribution to the total reaction rate and thus, requires further experimental efforts for verification. Our total reaction rate is slightly higher than previously available in the literature which could be served as an enhanced input to reaction network calculations for astrophysics.
A comparison of the  rate of $^{11}$B(n, $\gamma$)$^{12}$B reaction, which has implications not only in stellar plasma but could be a potential disruptor in proton-Boron fusion reactors, is done with those of the alpha and proton capture reaction available in the literature. We found that  $^{11}$B(p, $\alpha$)$^{8}$Be,  $^{11}$B($\alpha$, n)$^{14}$N and $^{11}$B(p, n)$^{14}$N reactions start dominating the $^{11}$B(n, $\gamma$)$^{12}$B at $T_9 \ge 0.2$, 1.35 and 2.8, respectively.  
Note that even excited states of $^{12}$B, which are possible neutron halo candidates \cite{Bely18PRC}, could contribute to the reaction rate as explored in Ref. \cite{Lin03PRC}, but we leave that for future due to lack of experimental evidences for the capture to its excited states.

\section*{Acknowledgements} [S] acknowledges SERB, DST, India for a Ramanujan Fellowship (RJF/2021/000176). [GS] would like to acknowledge PRISMA+ (Precision Physics, Fundamental Interactions and Structure of Matter) Cluster of Excellence, Johannes Gutenberg University Mainz.

\bibliography{gaganbiblio}

\backmatter

\section*{Declarations}

\begin{itemize}
\item Conflict of interest/Competing interests: The authors have no competing interests.
\item Authors' contributions: [S] and [GS] contributed towards ideation, calculations and manuscript preparation. [RC] supervised the work and reviewed critically. [MD] contributed towards intellectual content and manuscript preparation including revision.
\end{itemize}


\begin{flushleft}%
\bigskip\noindent

\bigskip\noindent
\bigskip\noindent
\bigskip\noindent
\end{flushleft}


\end{document}